\documentclass[letterpaper]{article} 
\usepackage{aaai23}  
\usepackage{times}  
\usepackage{helvet}  
\usepackage{courier}  
\usepackage[hyphens]{url}  
\usepackage{graphicx} 
\usepackage{natbib}  
\usepackage{caption} 
\usepackage{algorithm}
\usepackage{algorithmic}
\usepackage{amsmath}
\usepackage{bm}
\usepackage{subcaption}
\usepackage{booktabs}

\usepackage{newfloat}
\usepackage{listings}
\DeclareCaptionStyle{ruled}{labelfont=normalfont,labelsep=colon,strut=off} 
\floatstyle{ruled}
\newfloat{listing}{tb}{lst}{}
\floatname{listing}{Listing}
\title{Predicting Survival Time of Ball Bearings in the Presence of Censoring}

\author {
    Christian Marius Lillelund\thanks{Corresponding author.},
    Fernando Pannullo,
    Morten Opprud Jakobsen,\\
    Christian Fischer Pedersen
}
\affiliations {
    Department of Electrical and Computer Engineering, Aarhus University\\Finlandsgade 22, 8200 Aarhus N, Denmark\\
    \{cl,202102261,morten,cfp\}@ece.au.dk
}

\begin{document}

\maketitle

\begin{abstract}
Ball bearings find widespread use in various manufacturing and mechanical domains, and methods based on machine learning have been widely adopted in the field to monitor wear and spot defects before they lead to failures. Few studies, however, have addressed the problem of censored data, in which failure is not observed. In this paper, we propose a novel approach to predict the time to failure in ball bearings using survival analysis. First, we analyze bearing data in the frequency domain and annotate when a bearing fails by comparing the Kullback-Leibler divergence and the standard deviation between its break-in frequency bins and its break-out frequency bins. Second, we train several survival models to estimate the time to failure based on the annotated data and covariates extracted from the time domain, such as skewness, kurtosis and entropy. The models give a probabilistic prediction of risk over time and allow us to compare the survival function between groups of bearings. We demonstrate our approach on the XJTU and PRONOSTIA datasets. On XJTU, the best result is a 0.70 concordance-index and 0.21 integrated Brier score. On PRONOSTIA, the best is a 0.76 concordance-index and 0.19 integrated Brier score. Our work motivates further work on incorporating censored data in models for predictive maintenance.
\end{abstract}

\section{Introduction}

Ball bearings find extensive use in various rotary machines, but are susceptible to defects, such as contamination wear, poor lubrication, and improper mounting \cite{kim_prognostics_2017}. Monitoring wear in bearings plays an important role in predictive maintenance programs, and engineers have traditionally used frequency-based vibration analysis tools to assess the severity of emerging mechanical damages \cite{randall_vibration_2021, randall_rolling_2011}. Predictive maintenance should be done as soon as a bearings' operating characteristics start to deviate significantly from its normal operating state in order to avoid actual failure and eventual breakdown of the machinery. To this end, researchers have applied various machine learning (ML) algorithms, particularly neural networks, to train models that can predict the remaining useful life (RUL) of bearings \cite{guo_recurrent_2017, zheng_data-driven_2018, al_masry_remaining_2019, wang_hybrid_2020, wang_remaining_2022, xu_rul_2022}. This strategy has shown initial success and high predictive performance, but support for censored observations is generally overlooked in the field and only few studies have attempted to build models that support it \cite{widodo_machine_2011, widodo_application_2011, hochstein_survival_2013, wang_remaining_2022}. Given the rarity ball bearings failures, censored observations are common, and simply ignoring them can lead to a loss of efficiency and introduce estimation bias \cite{stepanova_censoring_2002}. Survival analysis is a type of regression that can leverage censored data \cite{gareth_introduction_2021}. Current works in bearing prognostics using survival analysis do not provide any evaluation of predictive accuracy or ranking performance \cite{wang_remaining_2022}, and the evaluated methods, such as the nonparametric Kaplan-Meier (KM) estimator \cite{kaplan_nonparametric_1958}, are often mainly descriptive, too simplistic or do not model the relationship between covariates and outcome \cite{widodo_application_2011}.

In this paper, we propose a novel approach for predictive maintenance of ball bearings using survival analysis (see Fig. \ref{fig:pipeline}). First, we process the raw bearing data in the frequency-domain to identify at which point in time a bearing starts to deviate significantly from its normal frequency characteristics. This happens well in advance of actual bearing failure, at a point where the faulty component can be readily replaced by the maintenance staff. This is our event detection algorithm, that we use to annotate bearings in two datasets (XJTU and PRONOSTIA) by the time this deviation occurs. The algorithm is a simple distance metric, and using it for forecasting is computationally expensive and prone to error. Therefore, we use the annotated data to train several survival models, that can predict the probability of failure as a progressive estimate over time, contrary to traditional RUL regression, that only offer a mere point estimate of the failure time. This method also enables us to quantify survival probabilities between groups of bearings by their time-domain features.

We compare our work to the method proposed by Xu et al. \cite{xu_rul_2022} to estimate the RUL of bearings in the PRONOSTIA dataset. However, since our methodology, the annotation algorithm we have used and the bearing data differ from \cite{xu_rul_2022} and others, we perform an indirect comparison. Source code is available at: \url{https://github.com/thecml/ball-bearing-survival}



\begin{figure}[!htbp]
	\centering 
	\includegraphics[width=0.45\textwidth]{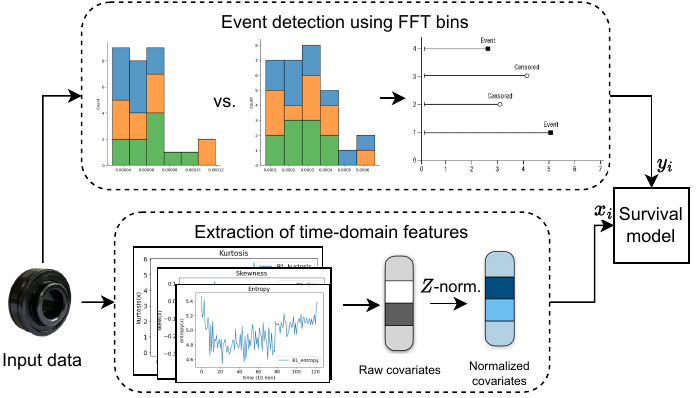}	
	\caption{Outline of the proposed solution. Raw data is used for event detection (top) and feature extraction (bottom) before training a survival model. Illustrations partly from \cite{nagpal_deep_2021}.} 
	\label{fig:pipeline}%
\end{figure}

\section{Fundamentals}
\subsection{Elements of Ball Bearings}
Ball bearings are used to carry rotating loads by separating elements in motion with two bearing races carried by steel or ceramic balls. Ball bearings can support either axial or radial loads, or a combination.




The health of a bearing can be tracked by monitoring the presence of defects in the bearing components. Bearing defects excite the bearing components and the resulting vibrations can be measured. By following the development of energy in selected frequency bands, ballpass frequency outer race (BPFO), ballpass frequency inner race (BPFI), fundamental train frequency (FTF), ball spin frequency (BSF), and shaft frequency (FS), each relating to a bearing's components, the bearing's health and remaining useful lifetime can be tracked \cite{wang_predictive_2017, randall_vibration_2021}. During infant bearing defects, the vibration signals will be weak and amplitude modulated with eigenfrequencies related to the bearing components, necessitating use of wideband Piezo or MEMS accelerometers (1--20 kHz) to capture any vibrations.

\subsection{Elements of Survival Analysis}
Survival analysis is a form of regression modeling that studies the time to an event, which can be partially observed (i.e., censored). Survival data contain the observed covariates, the time to event, and a label indicating whether the event occurred or was censored. We treat the survival time as discrete and limit the time horizon to a finite duration, denoted as $T = \{T_{0}, ..., T_{max}\}$, with $T_{max}$ representing the predefined maximum time horizon (e.g., 1 year). Within this framework, we consider bearing failure as the event of interest, and assume that exactly one such event will occur eventually for each observation (e.g., a bearing will eventually fail, but from only one cause). However, not all events of interest are always observed due to factors such as bearings being decommissioned before failure occurs or simply running problem-free after $T_{max}$, resulting in right-censored data.

Survival analysis models the probability that an event occurs at time $T$ later than $t$, which is denoted as the survival probability $S(t) = \Pr(T>t) = 1-\Pr(t\leq T)$. To estimate $S(t)$, we use the so-called hazard function, $h(t) = \lim_{\Delta t \rightarrow 0} \Pr(t<T\leq t+ \Delta t \vert T>t)/\Delta t$, which corresponds to the failure rate at an instant after time $t$, given survival past that time \cite{gareth_introduction_2021}. The relationship between the survival and hazard function is given by $h(t) = f(t)/S(t)$, where $f(t)$ is the probability density associated with $T$, $f(t) := \lim_{\Delta t \rightarrow 0} \Pr(t<T\leq t+\Delta t)/\Delta t $, which is the instantaneous rate of failure at time $t$. In this regard, $h(t)$ is the density of $T$ conditional on $T>t$, and the functions $S(t)$, $h(t)$, $f(t)$, are equivalent ways of describing the distribution of $T$,  which formalizes the intuition that higher values for $h(t)$ correspond to higher failure probabilities. In order to fit a regression model to survival times, Cox's Proportional Hazards (CoxPH) is a popular choice \cite{cox_regression_1972}. The model assumes a conditional individual hazard function of the form $h(t\vert \bm{x}_i) = h_0(t) \exp (f(\bm{\theta},\bm{x}_i))$, where $i$ denotes the $i$-th individual, $\bm{x}_i$ is a vector of $d$ covariates and $f(\bm{\theta},\bm{x}_i)$ is a linear function of the covariates.

\section{Methods and Materials}

\subsection{Datasets}

We consider two bearing datasets in this study, XJTU \cite{wang_hybrid_2020} and PRONOSTIA \cite{nectoux_pronostia_2012}. For both datasets, a single bearing under test is subjected to radial load during rotation, and vibrations are captured. The bearing is instrumented with two piezoelectric accelerometers, mounted horizontally and vertically, and sampled at 25.6 kHz. The resulting data are saved as raw comma-separated values. In both experiments, the bearings are run to failure under very high load (C/P$\leq$4) to accelerate degradation. This significantly increases the risk of initiating bearing defects but also comes with the risk of local flash heating and subsequent uncontrolled damage of a bearing.

\textbf{XJTU}: The dataset is generated from 15 deep grove ball bearings. Three types of tests are conducted with different loads and speeds, and five bearings are used in each test. In our experiments, we use five bearings with the following characteristics: a operating condition (C/P) of 1.1, a radial force of 12.0~kN and a rotating speed of 2100 RPM.

\textbf{PRONOSTIA}: The dataset is generated from 6 deep grove ball bearings. Three types of tests are conducted with different loads and speeds, where two bearings are used in each test. In our experiments, we use two bearings with the following characteristics: a C/P of 0.8, a radial force of 5.0~kN and a rotating speed of 1500 RPM.

\subsection{Feature extraction}

Features come from accelerometer sensors placed on the ball bearings at a angle of 90 degrees (referred to as the $X$-axis and $Y$-axis). Feature extraction starts by discretizing into bins a sequence of raw samples, i.e., $\bm{x} = (x)_{i=1}^N$. We then apply the expressions in Tab. \ref{tab:dist_tab} to each of these bins and obtain a total of twelve time-domain features for the whole lifetime of a bearing.

\begin{table}[!htbp]
\centering
\setlength\tabcolsep{1pt}
\def\arraystretch{1.5}
\begin{tabular}{|l|rl|}
\hline ~Absolute mean~ & $\bar{x}=$ & $\frac{1}{N} \sum_{i=1}^N\left|x_i\right|$ \\
\hline ~Standard deviation~ & $\sigma=$ & $ (\frac{1}{N} \sum_{i=1}^N (x_i-\bar{x})^2)^{1/2}$ \\
\hline ~Skewness~ & $SK=$ & $\frac{1}{N} \sum_{i=1}^N (x_i-\bar{x})^3 / \sigma^3$ \\ 
\hline ~Kurtosis~ & $K=$ & $\frac{1}{N} \sum_{i=1}^N (x_i-\bar{x})^4 / \sigma^4$ \\ 
\hline ~Entropy~ & $H =$ & $-\sum_{i=1}^N P(x_i) \log P(x_i)$~ \\
\hline ~Root-mean-square~ & ~$RMS=$ & $ (\frac{1}{N} \sum_{i=1}^N x_i^2)^{1/2}$ \\
\hline ~Max value~ & $max=$ & $\max(|\bm{x}|)$\\
\hline ~Peak-To-Peak~ &$P2P=$& $\max(|\bm{x}|) - \min(|\bm{x}|)$\\
\hline ~Crest factor~ &$CR=$& $max / RMS$\\
\hline ~Clearance factor~ &$CL=$ & $ x_p / (\frac{1}{N} \sum_{i=1}^N |x_i|^{1/2})^2$\\
\hline ~Shape factor~ & $SH=$ & $RMS / \bar{x}$\\
\hline ~Impulse~ & $IM=$ & $max / \bar{x}$\\
\hline
\end{tabular}
\caption{Features expressed in relation to a signal frame $\bm{x} = (x)_{i=1}^N$.}
\label{tab:dist_tab}
\end{table}

\subsection{Event detection}

\begin{figure}[!htbp]
	\centering 
	\includegraphics[width=0.9\columnwidth]{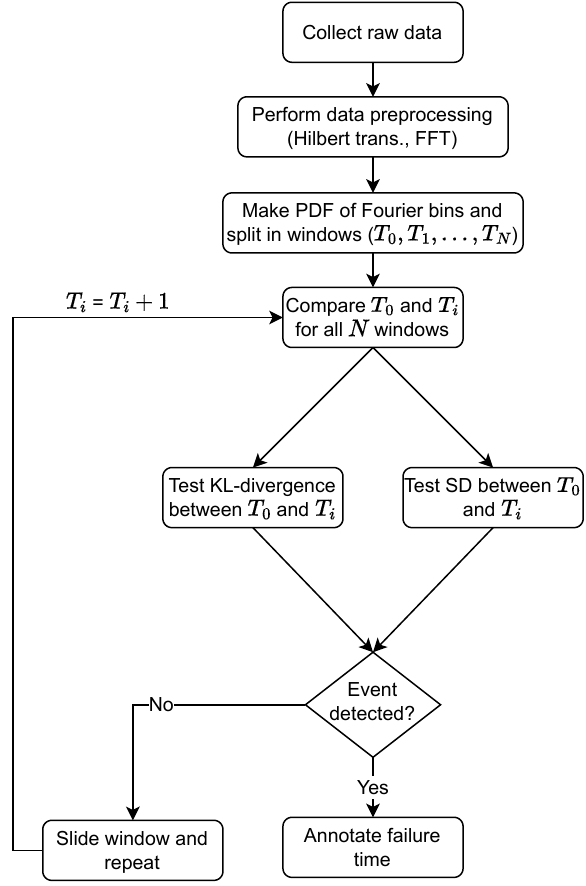}	
	\caption{Flowchart of proposed event detection algorithm. The entropy or deviation between the signal at time $T_0$ and time $T_{i}$ is continuously compared over the lifetime of the bearing.} 
	\label{fig:flowchart}%
\end{figure}

In this paper, we propose an event detection algorithm to label ball bearing data to be used in a survival model (see Fig. \ref{fig:flowchart}). Collecting the raw bearing data, performing the Hilbert transformation, and applying the fast Fourier transform (FFT), produces five main bins that make up a probability density function (PDF) for appropriate time windows. This allows us to observe different conditions of the bearing through time and enables us to identify the behavior and characteristic that describe a bearing in a good or bad state.

\begin{figure*}[!htbp]
    \centering
    \includegraphics[width=0.9\linewidth,trim={0, 0, 0, 0}]{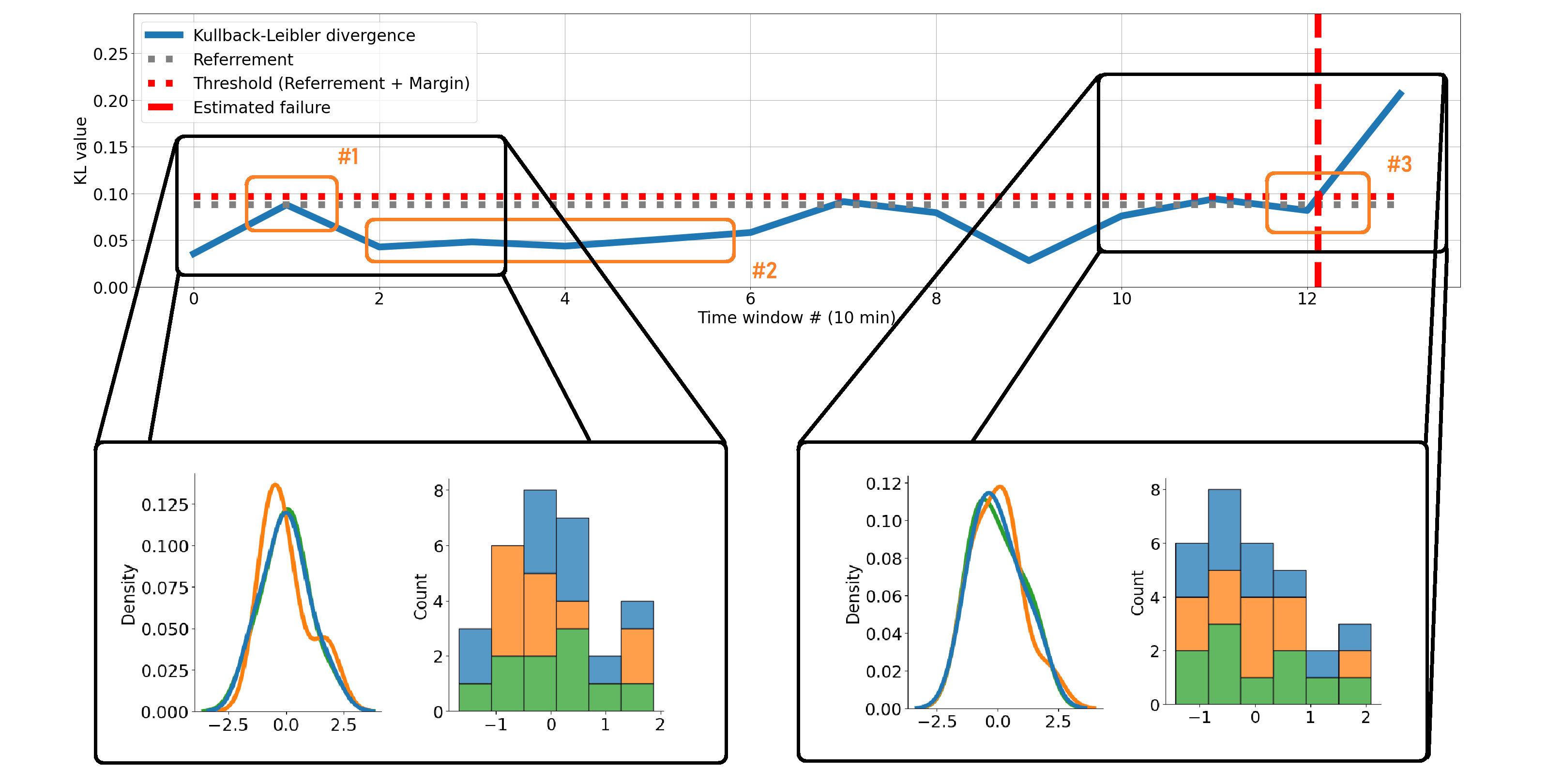}
    \caption{Top: Event detection based on KL divergence. The key point \#1 is the break-in period, \#2 is the plateau and \#3 is the break-out period. The solid blue line is the KL divergence. The dotted gray line is the maximum entropy reached during the break-in period. The dotted red line is the threshold given by the maximum entropy plus 10\%. Bottom left: FFT bins of the signal at break-in time. Bottom right: FFT bins of the signal at failure time.}
    \label{fig:event_detection}
\end{figure*}

After preprocessing, we apply the Kullback-Leibler divergence and standard deviation (SD) formulas in conjunction to detect changes in the PDF over time that possibly can identify the event of interest (see Fig. \ref{fig:flowchart}).  We consider two distributions: $T_{0}$ is the reference distribution and $T_{i}$ is one over a moving time window. In this context, the KL divergence is a statistical method that measures the difference between two probability distributions, denoted $D_{\mathrm{KL}}(P \| Q)$, as a measure of how one probability distribution $P$ is different from the probability distribution $Q$. This measures changes in entropy related to the distribution shape. We also compute the SD of $T_{0}$ and $T_{i}$, assuming they are Gaussian. By comparing the two PDFs' by SD and KL divergence from the break-in phase to the end of the bearing's lifetime, the progression of the PDF discrepancies are obtained. The KL is illustrated as a solid blue line in Fig. \ref{fig:event_detection}. The KL and SD progressions are similar, but not the same, because of changes to the underlying mechanics of the bearing throughout its lifetime. The KL progression line tends to reach a peak level (key point \#1 in Fig. \ref{fig:event_detection}), i.e., high entropy and toward diversification, which we use as a threshold (red dotted line in Fig. \ref{fig:event_detection}, the highest point of the blue line plus 10\%). Later, the threshold will be used to establish the event. After the break-in phase, the line reaches a plateau (key point \#2 in Fig. \ref{fig:event_detection}), i.e., low entropy and toward equalization, and then begins to rise again when the bearing starts to degrade, i.e., high entropy and towards diversification between the PDFs, and then failure occurs (key point \#3 in Fig. \ref{fig:event_detection}). We establish a KL-threshold and a SD-threshold and whichever of two are crossed last indicates an event.

\subsection{Data preprocessing}

The XJTU and PRONOSTIA datasets contain only few actual bearings, so the number of data points and events is low. To remedy this, we perform data augmentation to create synthesized versions of the two datasets. This process is done independently for each bearing. First, we up-sample (double) the data points by treating each bearing's $X$-axis and $Y$-axis as independent bearings. Second, we do Constrained Bootstrapping as in \cite{zelterman_bootstrap_1996}. Third, we divide the timeseries into 20 slices of the same size covering the entire lifetime, and for each slice assign a time to event or censoring, and the covariates at that time. This essentially re-samples a single event several times, while preserving the correlation between covariates and event time. We introduce a censoring rate of 20\% in both datasets by simulating that some bearings did not experience the event.

\subsection{Survival models}
\label{subsec:survival_models}

The following models are evaluated:

\textbf{Cox Proportional Hazards (CoxPH):} \cite{cox_regression_1972} is the most commonly used regression model for survival data. It assumes a conditional individual hazard function. The risk score is estimated as a linear function of covariates and parameters, found by maximizing the partial log-likelihood.

\textbf{Random Survival Forest (RSF):} \cite{ishwaran_random_2008} RSF is an ensemble of survival trees, where the data are recursively partitioned based on some splitting criterion, and similar data points based on the event of interest are put on the same node.

\textbf{CoxBoost:} \cite{hothorn_survival_2005} CoxBoost is an extension of traditional gradient boosting that supports survival data by minimizing the weighted empirical risk function as a least-squares problem.

\textbf{DeepSurv:} \cite{katzman_deepsurv_2018} DeepSurv is a neural network that uses a Cox likelihood function to compute a relative risk score, which quantifies the likelihood of experiencing an event.

\textbf{Deep Survival Machines (DSM):} \cite{nagpal_deep_2021} DSM is a neural network that estimates the conditional survival function as a mixture of primitive distributions, either Weibull or Log-Normal.

\textbf{Weibull AFT:} \cite{lee_statistical_2013} The Accelerated Failure Time (AFT) model estimates the survival function using the Weibull distribution. It is a generalization of the exponential distribution, but does not assume a constant hazard rate, which allows for a broader application.

\section{Experiments and Results}

\subsection{Setup}

For our empirical analyses, we use the openly available XJTU and PRONOSTIA datasets, which differ in the number of bearings, operating conditions, and the number of samples. After preprocessing, we apply a $z$-score data normalization to all covariates and do a train-test split based on the number of bearings: for XJTU, we select three bearings for training and two for test, and for PRONOSTIA, we select one bearing for train and one for test. For all datasets and models, we tune the following hyperparameters over ten iterations using 5-fold cross-validation:

\begin{itemize}
    \item CoxPH: Number of iterations and tolerance.
    \item RSF: Number of estimators, tree depth and split criterion.
    \item CoxBoost: Learning rate, tree depth, split criterion.
    \item DeepSurv: Iterations, learning rate and batch size.
    \item DSM: Iterations, learning rate and batch size.
    \item Weibull AFT: Penalizer coefficient.
\end{itemize}

Tuning is done solely on the training bearings. We use the hyperparameters leading to the highest average concordance index (Antolini's) on the validation folds to adjust the final models. Models are then evaluated on the test bearings.

\subsection{Results}
\label{subsec:results}

\begin{figure*}[!htbp]
  \centering
  \includegraphics[width=0.32\textwidth]{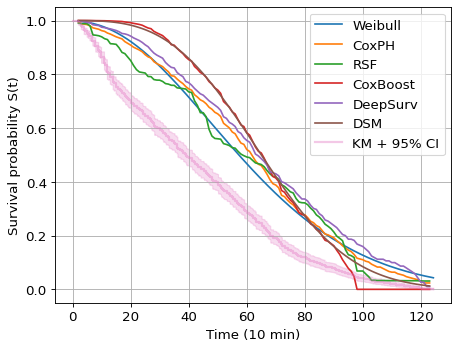}
  \includegraphics[width=0.32\textwidth]{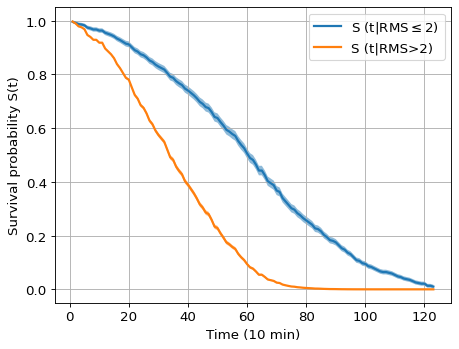}
  \includegraphics[width=0.335\linewidth]{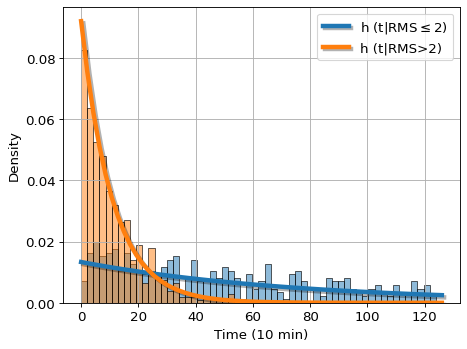}
\caption{Model inference on the XJTU dataset. Left \& center: survival function for different estimation approaches and for two separated groups based on RMS value (Solid lines: mean values; Shadow area: 95\% conf. int.). Right: normalized survival times (bins) and hazard functions (solid lines) of the two RMS groups.}
\label{fig:surv_plots}
\end{figure*}

Table \ref{tab:results} shows the predictive performance of the survival models. We report the training time, Harrell's concordance index (CI) \cite{harrell_multivariable_1996}, Antolini's time-dependent concordance index (CI\textsubscript{td}) \cite{antolini_time-dependent_2005} and the integrated Brier score (IBS) \cite{graf_assessment_1999}. Concerning the results on the XJTU dataset (see Tab. \ref{subtab:xjtu}), we observe that RSF, CoxBoost and DeepSurv perform the best in terms of ranking (CI) and DSM in terms of predictive accuracy. A CI of 0.70 means that 70 out of a 100 comparable pairs are ranked correctly. An IBS of 0.214 equals a 78.6\% accuracy in predicting the survival function over all available times. Concerning the results on the PRONOSTIA dataset (see Tab. \ref{subtab:pronostia}), DSM, DeepSurv and WeibullAFT performs the best.

To compare our work to SOTA, \cite{xu_rul_2022} report a mean 70\% accuracy (95\% CI: 0.58-0.87) on eleven PRONOSTIA bearings. In our work, DeepSurv \cite{katzman_deepsurv_2018} performs the best in terms of predicting the survival function, yielding a 0.197 integrated Brier score on the test set. This translates to a predictive accuracy of 80.3\%, however, the results are not directly comparable, as \cite{xu_rul_2022} use simple least-squares regression as opposed to survival regression, employ a different training/test split and only report the estimation error at the end of the bearing lifetime, instead of at available times.


Figure \ref{fig:surv_plots} shows inference results on the two XJTU test bearings. We see that all evaluated models provide consistent estimates of the survival function (leftmost panel). The Kaplan-Meier is used as a reference here and estimated on the full dataset (five bearings). In the central panel, we split the test data into two groups by the root mean square (RMS) covariate value, $RMS \leq 2$ and $RMS > 2$, and compute the mean survival curve of the two separately using a Cox proportional hazards model. This shows a clear separation in predicted survival probability, already from $t=10$, to convergence at $t=120$. The shadowed area represents a 95\% confidence interval, which is wider for the bearings in the first group. The plot presents an interesting application for predictive maintenance, as it can be used as decision support for maintenance personnel. In the rightmost panel, by following \cite{austin_generating_2012}, we computed 500 simulated survival times using a Cox model for the two RMS groups, and plotted their respective hazard function and survival time; the two hazard functions are quantitatively distinct, and the expected survival time for $RMS > 2$ is lower than for $RMS \leq 2$.

\section{Conclusion}

We have proposed survival analysis as a method to predict the risk of failure in ball bearings. First, we identify the point in time when a bearing starts to diverge significantly from its original frequency characteristic, which marks our event of interest, and second, we train several survival models to estimate the risk of failure given a set of covariates sampled from the time-domain. We find the application of such methods interesting in manufacturing and engineering, as they inherently support censored data and provide a probabilistic prediction and confidence estimation instead of mere point estimates, thus we encourage further work within our framework.

\section{Acknowledgments}
This work was supported by the PRECISE project under Grant Agreement No. AAL-2021-8-90-CP by the European AAL Association.



\begin{table}[!htbp]
\begin{subtable}{.45\textwidth}
\begin{center}
\caption{XJTU ($N=1000$, $C=20\%$, $d=12$).}
\label{subtab:xjtu}
\begin{tabular}{llllll} 
    {Model} & {T\textsubscript{train}} & {CI $\uparrow$} & {CI\textsubscript{td} $\uparrow$} & {IBS $\downarrow$} \\ \midrule
    CoxPH & \textbf{0.094} & 0.648 & 0.642 & 0.271 \\
    RSF & 0.988 & \textbf{0.702} & \textbf{0.769} & 0.423 \\
    CoxBoost & 2.791 & 0.59 & \textbf{0.769} & 0.261\\
    DeepSurv & 9.148 & 0.687 & 0.679 & 0.245\\
    DSM & 22.456 & 0.66 & 0.628 & \textbf{0.214}\\
    WeibullAFT & 0.483 & 0.664 & 0.675 & 0.216\\ \bottomrule
\end{tabular}
\end{center}
\end{subtable}%

\begin{subtable}{.45\textwidth}
\begin{center}
\caption{PRONOSTIA ($N=400$, $C=20\%$, $d=12$).}
\label{subtab:pronostia}
\begin{tabular}{llllll} 
    {Model} & {T\textsubscript{train}} & {CI $\uparrow$} & {CI\textsubscript{td} $\uparrow$} & {IBS $\downarrow$} \\ \midrule
    CoxPH & \textbf{0.016} & 0.626 & 0.625 & 0.34 \\
    RSF & 0.265 & 0.715 & 0.701 & 0.383 \\
    CoxBoost & 0.768 & 0.74 & 0.701 & 0.223\\
    DeepSurv & 2.392 & 0.745 & 0.705 & \textbf{0.197}\\
    DSM & 9.737 & 0.739 & \textbf{0.748} & 0.294\\
    WeibullAFT & 0.489 & \textbf{0.764} & 0.691 & 0.236\\ \bottomrule
\end{tabular}
\end{center}
\end{subtable}
\caption{Performance metrics on the test sets. $N$: total sample size, $C$: pct. of censored data, $d$: number of covariates.}
\label{tab:results}
\end{table}

\clearpage
\bibliography{references}

\end{document}